\documentclass[aps,prl,preprint, superscriptaddress,showpacs]{revtex4}

\usepackage[dvips]{graphicx,color}

\usepackage{bm}        
\usepackage{amssymb}   
\usepackage{amsmath} 
\def\pra#1#2#3{Phys.~Rev.~A~{\bf #1},\ #2\ (#3)}
\def\prl#1#2#3{Phys.~Rev.~Lett.~{\bf #1},\ #2\ (#3)}

\def\rmp#1#2#3{Rev.~Mod.~Phys.~{\bf #1},\ #2\ (#3)}

\def\k1{k_1}
\def\k2{k_2}
\def\q1{q_1}
\def\q2{q_2}

\def\({\left (}
\def\){\right )}
\def\[{\left [}
\def\]{\right ]}

\newcommand{\beq}{\begin{equation}}
\newcommand{\eeq}{\end{equation}}

\begin{document}
\date{\today}
\title{rf-field-induced Feshbach resonances}

\author{T. V. Tscherbul}
\affiliation{Harvard-MIT Center for Ultracold Atoms, Cambridge, Massachusetts 02138, USA}
\affiliation{ITAMP,
Harvard-Smithsonian Center for Astrophysics, Cambridge, Massachusetts 02138, USA}\email[]{tshcherb@cfa.harvard.edu}
\author{T. Calarco}
\affiliation{Institute of Quantum Information Processing, Ulm University, D-89081 Ulm, Germany}
\author{I. Lesanovsky}
\affiliation{School of Physics and Astronomy, The University of Nottingham, Nottingham NG7 2RD, UK}
\author{R.~V.~Krems}
\affiliation{Department of Chemistry, University of British Columbia, Vancouver BC, V6T 1Z1, Canada}
\author{A. Dalgarno}
\affiliation{Harvard-MIT Center for Ultracold Atoms, Cambridge, Massachusetts 02138, USA}
\affiliation{ITAMP, 
Harvard-Smithsonian Center for Astrophysics, Cambridge, Massachusetts 02138, USA} 
\author{J. Schmiedmayer}
\affiliation{Atominstitut, TU-Wien, Stadionallee 2, 1020 Vienna, Austria}

\begin{abstract}

A rigorous quantum theory of atomic collisions in the presence of radio frequency (rf) magnetic fields is developed and applied to elucidate the effects of combined dc and rf magnetic fields on ultracold collisions of Rb atoms. We show that rf fields can be used to induce Feshbach resonances, which can be tuned by varying the amplitude and frequency of the rf field. The rf-induced Feshbach resonances occur also in collisions of atoms in low-field-seeking states at moderate rf field strengths easily available in atom chip experiments, which opens up the world of tunable interactions to magnetically trappable atomic quantum gases.  
\end{abstract}

\maketitle

\clearpage
\newpage

Feshbach resonances induced by dc magnetic fields are a powerful tool to engineer interactions in ultracold quantum gases and they have led to significant advances in many-body physics, condensed matter physics, precision spectroscopy, quantum information processing, creating ultracold molecules, and facilitating evaporative cooling in optical dipole traps \cite{RMP,Bloch,SuperExchange,Molecules,Cs}. A number of related techniques make use of laser light \cite{OFR}, dc electric fields \cite{Roman}, or a combination of magnetic and optical fields \cite{Rempe} to shift the atomic energy levels in and out of resonance with those of a transient collision complex, thereby inducing resonant variation of the scattering length \cite{RMP}. 

Dressed state potentials induced by external radio frequency (rf) and microwave fields have remarkable properties. On atom chips \cite{AtomChip}, where substantial strength and gradients can be reached in the nearfield, complex potential landscapes (such as double-well and ring traps) can be created, which are ideal for matter wave interferometry \cite{Joerg1,Joerg2,MIT,MPQ} and exploring quantum many body physics \cite{Hofferberth}. Moreover, rf-dressed atoms are immune to spontaneous decay and it was suggested that low-frequency electromagnetic radiation can be used to control atomic collisions at ultralow temperatures  \cite{Joerg2}.

Early studies of atomic collisions in the presence of low-frequency electromagnetic radiation date back to the development of microwave traps for neutral atoms \cite{TrappingCs}, the studies of dipolar relaxation of spin-polarized hydrogen atoms in a microwave trap \cite{Agosta} and detailed analysis of rf-induced evaporative cooling \cite{Verhaar}. In the last few years, the creation of weakly bound KRb and Cr$_2$ molecules by rf-assisted association of ultracold atoms was demonstrated \cite{Zirbel,French}. Very recently,  multichannel quantum defect theory was applied to describe collisions of ultracold Rb atoms in rf fields \cite{Hanna}.

In this Rapid Communication, we develop a rigorous quantum theory of atomic collisions in the presence of rf fields. Expanding on the work of Agosta {\it et al.} \cite{Agosta}, we combine the coupled-channel expansion with the field-dressed formalism to describe the collision complex in an external rf field. This allows us to calculate the scattering length of ultracold Rb atoms as a function of the amplitude and frequency of the rf field. We demonstrate that collisions of ultracold atoms are sensitive to rf fields of moderate strength easily accessible in atom chip experiments. In particular, rf fields may be used to manipulate magnetic Feshbach resonances and to create new, rf-induced Feshbach resonances, whose positions and widths can be tuned by changing the rf field parameters. 

The Hamiltonian for two alkali-metal atoms A and B colliding in the presence of superimposed rf and dc magnetic fields may be written (in units of $\hbar$)
\begin{equation}\label{H}
\hat{H} = -\frac{1}{2\mu R}\frac{\partial^2}{\partial R^2} R + \frac{\hat{\ell}^2}{2\mu R^2} + \hat{V}_\text{sd}(R) + \hat{H}_\text{A} + \hat{H}_\text{B} +  \hat{H}_\text{rf},
\end{equation}
where $R$ is the interatomic distance, $\mu$ is the reduced mass of the collision complex, $\hat{\ell}$ is the orbital angular momentum for the collision, and $\hat{V}_\text{sd}(R)$ is the spin-dependent interaction potential between the atoms. We neglect the weak magnetic dipole-dipole interaction. The asymptotic Hamiltonian is $\hat{H}_\text{as}=\hat{H}_\text{A}+\hat{H}_\text{B}+\hat{H}_\text{rf}$. The terms $\hat{H}_\text{A}$ and $\hat{H}_\text{B}$ account for the interaction of individual atoms with a homogeneous rf field ${\bm B}(t)={\bm \hat{e}}B_\text{rf} \cos(\omega t)$
\begin{equation}\label{HA}
\hat{H}_\nu =  \textstyle{\frac{1}{2}}\gamma_\nu \hat{F}_\nu^2 + g_{F_\nu}\mu_0 {\bm B}\cdot \hat{F}_{\nu_z} +  \textstyle{\frac{1}{2}} \lambda_\nu (\hat{a}^\dag \hat{F}_{\nu_-} + \hat{a} \hat{F}_{\nu_+} + \text{H.c.}),
\end{equation}
where $\gamma_\nu$ is the hyperfine constant of atom $\nu$ ($\nu = \text{A},\,\text{B}$) with total angular momentum $\hat{F}_\nu$, $g_{F_\nu}$ is the magnetic $g$-factor, $\mu_0$ is the Bohr magneton, and ${\bm B}$ is the dc magnetic field vector, whose direction defines the quantization axis $z$. We assume that the rf field is polarized in the $x$-direction (${\hat{e}}={\bm \hat{x}}$). The Hamiltonian of the rf field $\hat{H}_\text{rf} =  \hbar\omega (\hat{a}^\dag\hat{a} - N)$ is expressed in a second-quantized form using the photon creation and annihilation operators $\hat{a}^\dag$ and $\hat{a}$, the average number of photons in the field $N$, and the rf frequency $\omega$. In Eq. (\ref{HA}), $\hat{F}_{\nu_\pm}$ are the atomic raising and lowering operators, and $\lambda_\nu =-\mu_0g_{F_\nu}B_\text{rf}/2\sqrt{N} $ is the coupling constant, which quantifies the strength of the atom-field interaction. In the rotating-wave approximation, the Hermitian conjugate terms in Eq. (\ref{HA}) are neglected in calculating the matrix elements between the states within the $F=2$ manifold (and vice versa for the $F=1$ manifold).

We consider ultracold collisions of two identical $^{87}$Rb atoms in superimposed dc and rf magnetic fields. Following Agosta {\it et al.} \cite{Agosta}, we introduce a symmetrized basis
\begin{equation}\label{basis}
\frac{1} { [2(1+\delta_{\tau_\text{A} \tau_\text{B}})]^{1/2} }  \left[ |\tau_\text{A} \tau_\text{B}\rangle +\eta(-)^\ell |\tau_\text{B} \tau_\text{A}\rangle \right] |N+n\rangle |\ell m_\ell\rangle,
\end{equation}
where $|\tau_\nu\rangle$ are the hyperfine states $|F_\nu m_{F_\nu}\rangle$, $|N+n\rangle$ are the photon number states, $\eta$ accounts for exchange symmetry ($\eta=+1$ for two identical $^{87}$Rb atoms), and $|\ell m_\ell\rangle$ are the partial wave states.  Diagonalization of $\hat{H}_\text{as}$ in the basis (\ref{basis}) yields the energy levels of two non-interacting atoms dressed by the rf field. The field-dressed states occur in manifolds separated by multiples of the photon energy $\hbar \omega$ \cite{Book}. Throughout this work, we consider rf frequencies in the range $\omega/2\pi=18$-40 MHz and moderate dc fields $B<10$ G, which corresponds to the weak coupling regime, where the energy separation between different photon manifolds is larger than the coupling between the manifolds \cite{CCT}.  We use the eigenfunctions of the asymptotic Hamiltonian to expand the wave function of the collision complex. Inserting this expansion into the Schr{\"o}dinger equation leads to a system of close-coupled (CC) differential equations yielding $S$-matrix elements and transition probabilities between the field-dressed states. To solve the CC equations, we evaluate the matrix elements of the Hamiltonian (\ref{H}) in the field-dressed basis analytically \cite{NJP} using the most accurate potentials for the singlet and triplet states of Rb$_2$ \cite{Zhiying}. Seven photon number states are included in the basis set (\ref{basis}) resulting in 252 coupled equations, which are integrated on a grid of $R$ from 3 to 300 $a_0$ with a step size of $5\times 10^{-3}$ $a_0$. This procedure produces fully converged results at a collision energy of 1~$\mu$K. To evaluate the bound states of the Rb$_2$ molecule, we use the asymptotic bound-state model (ABM) \cite{ABM} properly extended to include the interactions with rf fields.

\begin{figure}[t]
	\centering
    \includegraphics[width=0.55\columnwidth]{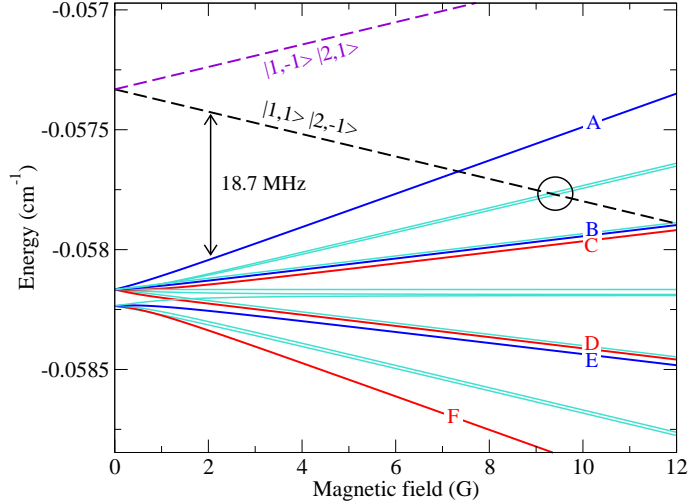}	
	\renewcommand{\figurename}{Fig.}
	\caption{Near-threshold energy levels of Rb$_2$ vs magnetic field. The energies of two separated atoms are shown by dashed lines and labeled in the graph. The red (blue) lines show the molecular levels A-F (in order of increasing binding energy) with $m_F = m_{F_\text{A}} + m_{F_\text{B}}=-1$ (+1). The light blue lines correspond to the molecular states with $m_F\ne \pm 1$. The zero of energy corresponds to two infinitely separated Rb atoms in the absence of hyperfine structure and external fields. }
\end{figure}

Figure 1 shows the magnetic field dependence of near-threshold energy levels of the Rb$_2$ molecule in the absence of an rf field \cite{ABM}. At certain magnetic fields, the asymptotic two-atom scattering states become nearly degenerate with molecular states, giving rise to Feshbach resonances \cite{RMP}.
We focus on a mixed-spin resonance that has been detected experimentally in an optically trapped mixture of ultracold Rb atoms in the $|1,1\rangle \otimes |2,-1\rangle$ initial state  \cite{FRexp1,FRexp2}. This resonance corresponds to the symmetry-allowed crossing encircled in Fig. 1. Figure 2(a) shows the real and imaginary parts of the scattering length $a=\alpha-i\beta$ for two Rb atoms as functions of dc magnetic field for different rf field amplitudes. The elastic and inelastic cross sections can be evaluated from $a$ as $\sigma_\text{el}=8\pi (\alpha^2 + \beta^2)$ and $\sigma_\text{inel} = 8\pi\beta/k_i$ \cite{RMP}.  At zero rf field, the resonance occurs at $B=9.15$ G, in good agreement with the experimental results of 9.09$\pm 0.01$ G \cite{FRexp1} and $9.12(9)$ G \cite{FRexp2}. 
As shown in Fig. 2, an external rf field shifts the resonance to higher magnetic fields without affecting its width. The magnitude of the shift increases quadratically with increasing rf amplitude. 
An expanded view of the encircled area in Fig. 1 presented in Fig. 2(c) illustrates that off-resonant rf fields modify both atomic thresholds and molecular bound states, thereby altering the positions of avoided crossings and Feshbach resonances. The shifts arise due to virtual exchange of rf photons between different atomic hyperfine states, which modifies the atomic $g$-factors \cite{CCT}. The ABM overestimates the resonance position by 0.15 G, but accurately reproduces the rf-induced resonance shift. 


\begin{figure}[t]
	\centering
      \includegraphics[width=0.45\textwidth]{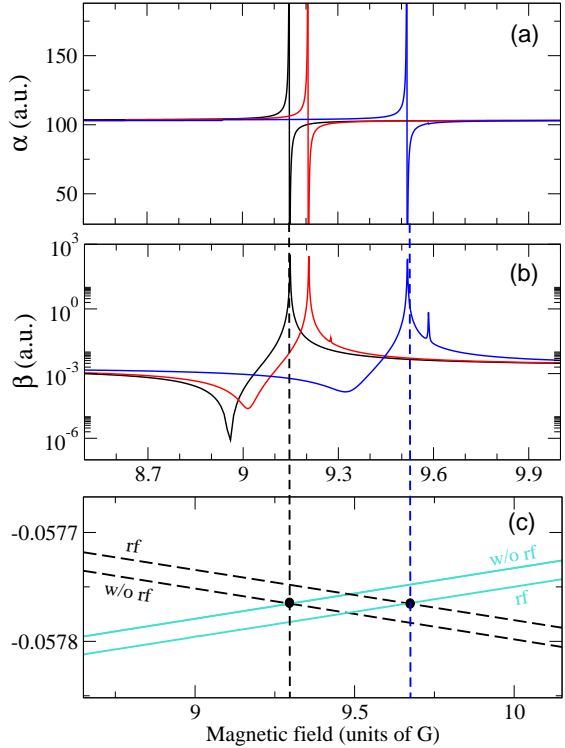}
	\renewcommand{\figurename}{Fig.}
	\caption{Magnetic field dependence of the real (a) and imaginary (b) parts of the scattering length for $s$-wave collisions of $^{87}$Rb atoms initially in the $|1,1\rangle\otimes |2,-1\rangle$ state at different rf field strengths: zero (left), $B_\text{rf}=4$ G (middle), and $B_\text{rf}=10$ G (right). (c) A zoom into the encircled area in Fig.~1 showing atomic thresholds (dashed lines) and molecular bound states (full lines) at zero rf field and $B_\text{rf}=10$~G. $\omega/2\pi = 30$ MHz. }
\end{figure}

\begin{figure}[t]
	\centering
     \includegraphics[width=0.5\textwidth]{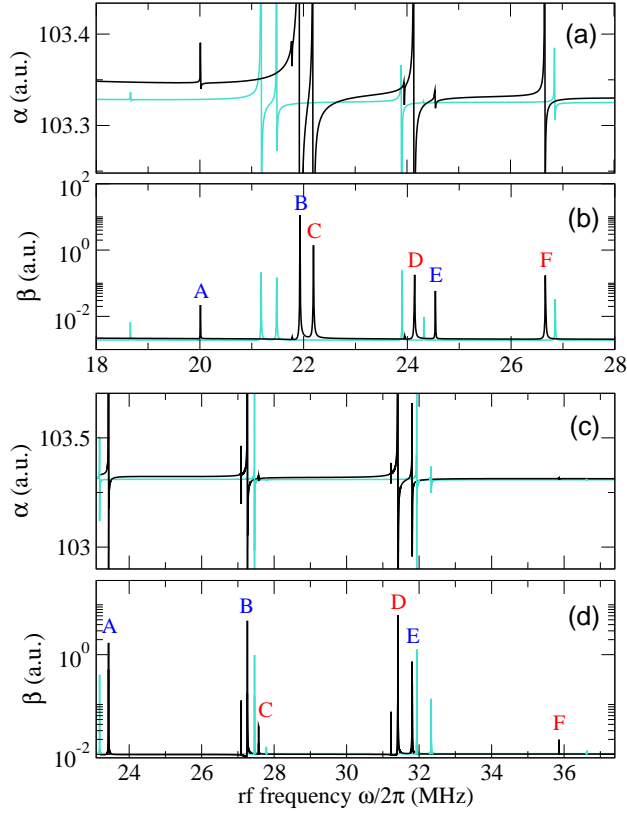}
	\renewcommand{\figurename}{Fig.}
	\caption{The real (a) and imaginary (b) parts of the scattering length for $s$-wave collisions of $^{87}$Rb atoms initially in the $|1,1\rangle\otimes |2,-1\rangle$ state as functions of the rf frequency at $B=2$ G. (c), (d) Same as in (a), (b) but for the $|1,-1\rangle\otimes |2,1\rangle$ initial state and $B=3.23$ G. The rf amplitude is 4~G [blue (light grey) lines] and 10~G (black lines). The rf-induced resonances are labeled after the bound states A-F in Fig. 1.}
\end{figure}

An interesting possibility suggested by recent experimental work \cite{Zirbel} is to use rf fields to directly couple a two-atom scattering state with a bound molecular state. This coupling would lead to a Feshbach resonance whose properties depend on the parameters of the rf field. 
As shown in Fig. 1, six of 15 low-lying molecular states can be accessed from the continuum by scanning the rf frequency from 18 to 24 MHz. The other 11 states remain uncoupled from the initial state $|1,1\rangle \otimes |2,-1\rangle$ due to the selection rules $\Delta n=\pm 1$ and $\Delta m_F=\mp 1$ imposed by Eq. (\ref{HA}).
Figure~3(a)-(b) show the rf frequency dependence of the cross sections for two Rb atoms in the $|1,1\rangle \otimes |2,-1\rangle$ state at a magnetic field of 2 G. The scattering length exhibits a number of sharp peaks at rf frequencies corresponding to the bound levels A-F in Fig.~1. The widths of the resonances increase with increasing rf amplitude. As shown in Fig. 3(c)-(d), rf-induced Feshbach resonances can also occur in collisions of atoms initially in the {\it low-field-seeking state} $|1,-1\rangle\otimes|2,1\rangle$. 


In order to understand the mechanism of rf-induced Feshbach resonances, we use a three-state model developed by Bohn and Julienne \cite{BJ} and illustrated in Fig. 4. The rf field couples the incident channel $|i,N\rangle$ to the field-dressed bound state $|b,N+1\rangle$ with energy $\epsilon_b+\hbar\omega$, where $\epsilon_b$ is the binding energy of the molecular state in the absence of an rf field. The bound state $|b,N+1\rangle$ is coupled to an open channel $|a,N+1\rangle$ by the spin-dependent interaction potential (\ref{H}). Introducing the matrix elements $\gamma_{ib}=2\pi|\langle i,{N}| \hat{H}_\text{A} + \hat{H}_\text{B}|b,N+1\rangle|^2$ and $\gamma_{ba}=2\pi|\langle b,N+1 |\hat{V}_\text{sd}(R)|a, N+1 \rangle|^2$, we obtain for the real and imaginary parts of the scattering length \cite{BJ}
\begin{equation}\label{alpha}\notag
\alpha = \alpha_\text{bg} + \frac{1}{k_i} \frac{ {\textstyle \frac{1}{2}}\gamma_{ib}\hbar [\omega-\omega_b - \delta \omega(B_\text{rf})] }{ \hbar^2 [\omega-\omega_b - \delta\omega_b(B_\text{rf})]^2 + \gamma_{ba}^2/4},
\end{equation}
\begin{equation}\label{sl}
\beta = \frac{1}{k_i} \frac{ {\textstyle \frac{1}{4}}\gamma_{ib}\gamma_{ba}}{ \hbar^2 [\omega-\omega_b - \delta\omega_b(B_\text{rf})]^2 + \gamma_{ba}^2/4},
\end{equation}
where $k_i$ is the wave vector for the incident channel, $\alpha_\text{bg}$ is the background scattering length, $\omega_b$ is the resonant rf frequency ($\hbar\omega_b = \epsilon_b$), and $\delta\omega(B_\text{rf})$ is the rf-induced level shift. The expressions (\ref{sl}) are similar to those encountered in the theory of optical Feshbach resonances \cite{RMP,BJ} with one  important difference: the loss from the bound state $|b\rangle$ is induced by the spin-dependent interaction potential (\ref{H}) rather than spontaneous emission \cite{OFR,BJ}. Since $\gamma_{ib}\sim B_\text{rf}^2$, the widths of rf-induced resonances increase quadratically with $B_\text{rf}$. The resonant frequencies given by Eqs. (\ref{sl}) are shifted from those corresponding to bare molecular states by the amount $\delta\omega(B_\text{rf})\sim B_\text{rf}^2$ \cite{BJ}. Because the rf-induced coupling is weak and $\alpha_\text{bg}$ is large, the real part of the scattering length does not vary appreciably in the vicinity of the resonances, increasing from 103.7 to 106.5 $a_0$ near the resonance D as the rf frequency is scanned from 31.405 to 31.413 MHz. The inelastic loss rate at the resonance is $\sim$10$^{-12}$ cm$^{3}$/s. From Eqs.  (\ref{sl}), it is clear that $\alpha$ can be controlled more efficiently in atomic systems with smaller $\alpha_\text{bg}$. Alternatively, $\alpha_\text{bg}$ can be reduced using magnetic \cite{RMP} or electric \cite{Roman} Feshbach resonances.

As follows from Eqs. (\ref{sl}), $\alpha$ and $\beta$ peak at different rf detunings: $\Delta_\alpha = \pm \gamma_{ba}/2$ and $\Delta_\beta = 0$, with peak heights $\sim\gamma_{ib} / \gamma_{ba}$. Therefore, to induce an appreciable variation of $\alpha$ while keeping inelastic losses to a minimum, it is necessary to operate at large rf detunings $\Delta \gg \gamma_{ba}$, where $\alpha-\alpha_\text{bg}\sim \gamma_{ib}/\Delta$ and $\beta\sim \gamma_{ib}\gamma_{ba}/\Delta^{2}$. In this regime, the real part of the scattering length is independent of $\gamma_{ba}$. For practical applications, however, it may be advantageous to minimize inelastic losses quantified by the parameter $\gamma_{ba}$, so that the condition $\Delta \gg \gamma_{ba}$ is satisfied at smaller $\Delta$, where $\alpha-\alpha_\text{bg}$ is most sensitive to $\Delta$. Indeed, as shown in Fig.~3, $\beta$ is never too large except at very small detunings, which indicates that the regime $\Delta \gg \gamma_{ba}$ is within reach.




\begin{figure}[t]
	\centering
	\includegraphics[width=0.550\textwidth]{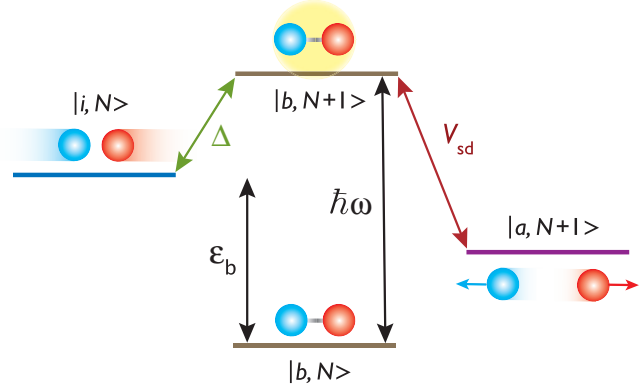}
	\renewcommand{\figurename}{Fig.}
	\caption{A schematic illustration of the rf-induced Feshbach resonance. The rf field induces coupling between the initial two-atom dressed state $|i,N\rangle$ and a bound molecular state $|b,N+1\rangle$, which is coupled with a low-lying open channel $|a,N+1\rangle$ via the spin-dependent interaction potential. $\Delta = \hbar\omega - \epsilon_b$ is the detuning from resonance.}
\end{figure}

In summary, we have developed a rigorous quantum theory of atomic collisions in the presence of rf fields, and applied it to explore the effect of rf fields on $s$-wave collisions of Rb atoms. 
Our results show that rf fields generated in experiments with atom chips can be used as an efficient tool to control collisions of ultracold atoms, which has a number of advantages over the methods that use dc magnetic fields.
In particular, shifting magnetic Feshbach resonances with rf fields may be used to selectively tune scattering lengths in ultracold atomic mixtures, thereby enabling the preparation of exotic many-body states and novel quantum phases \cite{Ueda}. As illustrated in Fig. 2(a), applying an rf field with $B_\text{rf}=4$ G at $B= 9.206$ G results in a three-fold enhancement of the real part of the scattering length. The calculated inelastic decay rates do not exceed $\sim$10$^{-10}$ cm$^3$/s, so elastic collisions of Rb atoms can be efficiently manipulated on timescales of hundreds of microseconds for typical atom densities of $10^{13}$~cm$^{-3}$. 
Further advantages of rf control include the possibility of tuning the rf frequency with extremely high precision, which may allow for ultra-sensitive measurements of molecular binding energies and scattering properties. Since rf fields can be switched in a very fast and controllable way \cite{Joerg1}, the rf-shifted Feshbach resonances offer the possibility of tuning the interaction between the atoms on a much shorter timescale than is possible using dc magnetic fields alone. The fast manipulation of ultracold atomic gases may be exploited to enhance the fidelity of quantum logic gates based on collisional interactions. In particular, a controlled phase gate, equivalent to a quantum CNOT, can be realized with nearly perfect fidelity by ``activating'' the atom-atom interaction, via rf fields, for the time needed for the system to acquire a $\pi$ phase nonlinearity among the two-atom logical basis (i.e. hyperfine) states, see e.g. \cite{Calarco2004}. The ability to control spin-dependent interactions is also important in the context of simulating highly correlated many-body systems with ultracold atoms in optical lattices \cite{SuperExchange}. On atom chips, near-field rf control would add the capability of addressing individual pairs of atoms, an essential feature for universal quantum computation that is known to be difficult to realize with optical lattices.

We thank J. Thywissen, P. Julienne, and T. Hanna for discussions and E. Tiesinga for providing the Rb$_2$ potentials. This work was supported by NSERC of Canada, the Austrian Science Foundation grant FWF-Z 118-N16, the EC (projects SCALA, CHIMONO), and the Chemical Science, Geoscience, and Bioscience Division of the Office of Basic Energy Science, Office of Science, U.S. Department of Energy and NSF grants to the Harvard-MIT CUA and ITAMP at Harvard University and Smithsonian Astrophysical Observatory.



\end{document}